\documentclass[12pt]{article}		%use with showkeys or to print one sided
\usepackage[super,sort,comma]{natbib}
\usepackage[hidelinks]{hyperref}
\usepackage{caption}
\usepackage{booktabs}
\usepackage{subcaption}
\usepackage{tabularx}
\usepackage{setspace} % for spacing control
\usepackage{booktabs}
\usepackage{booktabs}
\usepackage{hhline}
\usepackage{fancyhdr}		%Gives headers and footers defined below
\usepackage{amsmath}

\usepackage[section]{placeins} 
\usepackage{graphicx}
\usepackage{wrapfig}
\makeatletter \renewcommand\@biblabel[1]{$^{#1}$} \makeatother
\newcommand{\cen}[1]{\begin{center} #1 \end{center}}

\usepackage[mathlines]{lineno}
\usepackage{hyperref}
\hypersetup{ colorlinks,
    citecolor=blue,
    filecolor=blue,
    linkcolor=blue,
    urlcolor=blue
}
\usepackage{xcolor}
\definecolor{gray}{rgb}{0.6,0.6,0.6}
\definecolor{red}{rgb}{0.85,0,0}
\definecolor{green}{rgb}{0,0.85,0}
\definecolor{blue}{rgb}{0,0,0.85}
\definecolor{beige}{rgb}{0.92,0.87,0.78}
\usepackage[all]{hypcap}    %causes link to figures to go to figure, not caption
\begin{document}
\begin{center}
    \cen{\sf {\Large {\bfseries Robust External-Beam Calibration of Plastic Scintillation Detectors for In-Vivo Dosimetry in HDR Brachytherapy}}}
\vspace*{5mm}
{Chahrazed Ghannoudi$^{1,2}$, Marie-Claude Lavallée$^{1,2}$, Benjamin Côté$^{3}$, Luc Beaulieu\textsuperscript{\dag}$^{1,2}$ }\\
\vspace*{5mm}
{
\small%
    $^1$Département de physique, de génie physique et d’optique et Centre de recherche sur le cancer, Université Laval, Québec, Canada\\
    $^2$Département de radio-oncologie et Axe Oncologie du CRCHU de Québec, CHU de Québec Université Laval, Québec, QC, Canada\\
    $^3$Medscint Inc., Québec, QC, Canada\\
\vspace{2mm}
}
\pagenumbering{roman}
\setcounter{page}{1}
\pagestyle{plain}
\small
\textsuperscript{\dag}Author to whom correspondence should be addressed. email: luc.beaulieu@phy.ulaval.ca
\end{center}
\begin{abstract}
\noindent {\bf Objective.} HDR brachytherapy is a widely adopted modality for cancer treatment. However, it is not free from error and uncertainty. In-vivo dosimetry (IVD) is the only technique that confirms correct dose delivery. This study details and validates a calibration method for Plastic Scintillation Detector (PSD), bypassing dose gradient and positioning issues in brachytherapy calibration. {\bf Approach.} The PRB-0057 PSD (Medscint, Québec, Canada), a $1\times1$ mm scintillating fiber coupled to a $20$ m Eska GH-4001 clear optical fiber (Mitsubishi Rayon, Japan) of the same diameter, was calibrated and connected to the Hyperscint-RP200 research platform for optical signal collection. Hyperspectral calibration was performed at a LINAC with a 6 MV beam, enabling the independent stem effect removal. For validation, brachytherapy measurements with a $S_k=29447$ U Iridium-192 Flexisource (Elekta Brachy, Veenendaal, The Netherlands) were performed in a motorized IBA-Blue-Phantom\textsuperscript{2} water tank. Dose rates were measured at $10$ Hz along the source’s vertical z-axis at a fixed transverse distance of $1.2\pm0.05$ cm in $0.2$ cm steps. Calibration accuracy was evaluated using relative differences (RD) between measured and TG-43U1 dose rates, converted to positional errors. A detailed uncertainty budget was established to the measurement setup. {\bf Main results.} Comparison agreed with RDs around $2.5\%$ at $1.2$ cm, corresponding to positional uncertainties of \(<0.15\) mm. At greater distances, up to $8$ cm, RDs increase to about $5\%$, corresponding to positional uncertainties up to 3 mm, mainly due to reduced light-yield. Uncertainties found to depend on the source-detector distance, ranging from $3.81\%$ to $6.39\%~(k=1)$ over the range of explored distances. {\bf Significance.} Results confirm the effectiveness of a 6 MV external beam PSD calibration to be used in time-resolved IVD. Uncertainties close to the source are consistent with the Afterloader and IBA motorized unit reproducibility and are mainly dominated by reduced detector sensitivity at larger distances. Our study further underlined the intrinsic limitation of IVD in the face of known uncertainties.
\end{abstract}
\setlength{\baselineskip}{0.7cm}	
\pagenumbering{arabic}
\setcounter{page}{1}
\pagestyle{fancy}
\clearpage
\section{Introduction}
In High-Dose-Rate (HDR) brachytherapy (BT), a sealed source is manually or automatically put into or close to the target. This modality is widely adopted for numerous cancer sites, especially prostate cancer. However, despite advancements in treatment techniques, HDR BT is not free from potential errors, mainly those arising from manual procedures and the high dose gradient around the source \cite{Valentin2005, Kirisits2014, DeWerd2011, Mitch2007}. For example, the latter could induce dose uncertainties due to inaccuracies in source positioning, from applicator reconstruction to imaging limitations. More precisely for prostate BT, offsets of 2–6 mm in three or more needles result in $>10\%$ changes in DVH metrics, as demonstrated in a transrectal ultrasound-based real-time HDR prostate brachytherapy treatments study \cite{Poder2019}. Also, Buus et al. have shown that overall implant migrations in an HDR prostate treatment of 3 mm and 5 mm reduced prostate doses by 5\% and 10\%, respectively \cite{Buus2018}. Thus, HDR BT process includes several critical steps, starting from the calibration of the radioactive source, planning of the irradiation positions, manual insertion of catheters, and their reconstruction, each of which carries potential sources of error applicable to numerous treatment sites \cite{Kirisits2014}. The major challenge, therefore, lies in maintaining optimal accuracy throughout the entire treatment workflow.

In order to ensure precise treatment delivery, in-vivo dosimetry (IVD) has been emphasized by a recent ESTRO report, highlighting the importance of real-time treatment verification and outlining the requirements and future directions for IVD in brachytherapy \cite{Fonseca2020}. Previous studies have shown the potential of Plastic Scintillation Detectors (PSDs) as effective dosimeters for HDR brachytherapy \cite{Lambert2007, Tanderup2013}. Scintillation is a phenomenon in which a material emits light when excited by charged particles through a Förster-type resonance energy transfer process \cite{Beaulieu2016}. These materials are characterized by physical properties similar to those of human tissues, a linear response with the deposited dose, and the ability to provide real-time readings. Additionally, plastic scintillation dosimeters are proven to be energy independent \cite{Beddar2006}, making correction factors negligible compared to other available IVD dosimeters (i.e. diamonds, MOSFETs). These characteristics make them particularly well suited for clinical \textit{in-vivo} use, in addition to their small size ($\leq$ 1 mm). Inserted in the needles or in the rectal catheter PSDs would precisely track the source and report the radiation delivered to a point, as the light yield is linear to dose. However, PSDs are subject to light contamination, known as the \textit{stem effect}, caused by Cherenkov and fluorescence light emitted in the optical fiber, which is used to guide the scintillation signal to the light detector \cite{Beddar2016, Therriault2013}.  This stem effect must be corrected to ensure accurate dosimetry. Techniques such as spectral subtraction \cite{Beddar1992}, temporal separation \cite{Beddar2003}, and spectral filtering \cite{Archambault2006} have been developed to correct this effect in the output signal. 

In full brachytherapy conditions, the detector should be precisely positioned according to the source. Small deviations in the high dose gradient regions during experimental calibration would lead to inaccurate stem effect estimation and scintillation light separation, and thus imprecise dose calibration. Georgi et al.\ \cite{Georgi2024} recently highlighted these limitations and showed that traditional single-point calibration are less accurate  in full BT conditions without precise control of detector geometry/coordinates. They proposed an in-clinic calibration routine using multiple dwell positions to compensate for positional uncertainties, achieving an overall accuracy below 5\%. This work confirms the importance of adopting calibration techniques that remain robust in steep dose-gradient fields by recommending calibration with other sources. Thus, this study proposes a simple and robust calibration method for a single point plastic scintillation detector using a 6 MV beam from a linear accelerator, bypassing the issue of dose gradient and dosimeter positioning encountered in HDR brachytherapy. Validated in BT conditions, we further demonstrate that the stem effect is correctly removed, even in extreme brachytherapy geometry, and extract a detailed uncertainty budget for our dosimetry system and experimental set-up.
\section{Methods}
\subsection{Dosimetry system}
\begin{wrapfigure}{r}{0.42\linewidth}
    \centering
    \vspace{-0pt} 
    \includegraphics[width=0.55\linewidth]{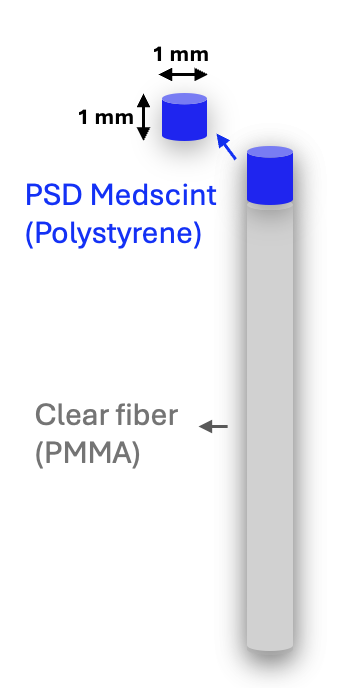}
    \caption{Schematic of the single-point plastic scintillation detector (PSD) from Medscint (QC, Canada), composed of a 1~mm diameter, 1~mm long polystyrene scintillator optically coupled to a clear PMMA optical fiber.}
    \label{fig:PSDConfig}
    \vspace{-0pt} 
\end{wrapfigure}

The FLEX Series PSD detector (PRB-0057, Medscint, Québec, Canada), consisting of a 1 mm diameter and 1 mm long (1 mm$^3$) cylindrical scintillating fiber, was coupled to a 20 m Eska GH-4001 clear optical fiber (Mitsubishi Rayon, Tokyo, Japan) of the same diameter to guide the optical light to the photodetector (Figure  \ref{fig:PSDConfig}). The fiber was connected to the Hyperscint-RP200 (HS RP200) spectrometer, the HyperDose RP v1.5.3 a multi-channel scintillation dosimetry platform from Medscint Inc, collecting the optical signal using one channel for this study.
\subsection{Detector calibration}
\begin{figure}[h!]
    \centering
    \includegraphics[width=1\linewidth]{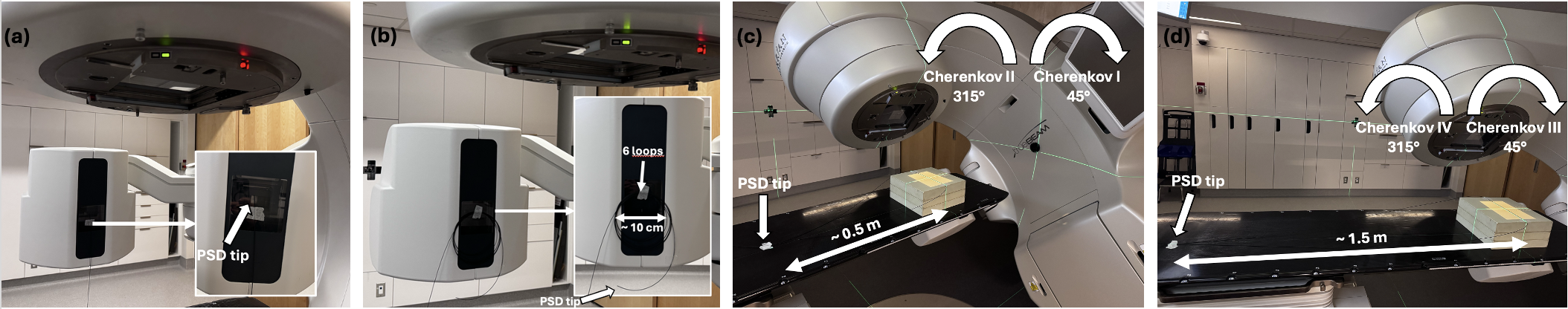}
    \caption{Spectral calibration setup for the plastic scintillation detector (PSD) using a TrueBeam linear accelerator. \textbf{(a)} PSD positioned at isocenter with the tip aligned to the Linac kV source for scintillation spectra extraction. \textbf{(b)} Fluorescence spectra extraction while 6 loops of the clear fiber of around $10~cm $ of diameter is placed at the Linac kV source. \textbf{(c)} Acquisition of reference spectra for Cherenkov contribution I and II  by irradiating the optical fiber at gantry angles $45^{\circ}$ and $315^{\circ}$, respectively, with the PSD tip placed outside the primary field at around $0.5~cm$. \textbf{(d)} Measurement geometry at extended source-to-detector distances ($\sim1 ~m$) for Cherenkov contributions III and IV. Both \textbf{(c)} and \textbf{(d)} setups used a 6 MV beam and a $15\times15~cm$ field size. All spectra acquisitions were performed at 1 Hz.\\}
    \label{fig:SpecCalib}
\end{figure}

\begin{figure}[h!]
    \centering
    \includegraphics[width=0.5\linewidth]{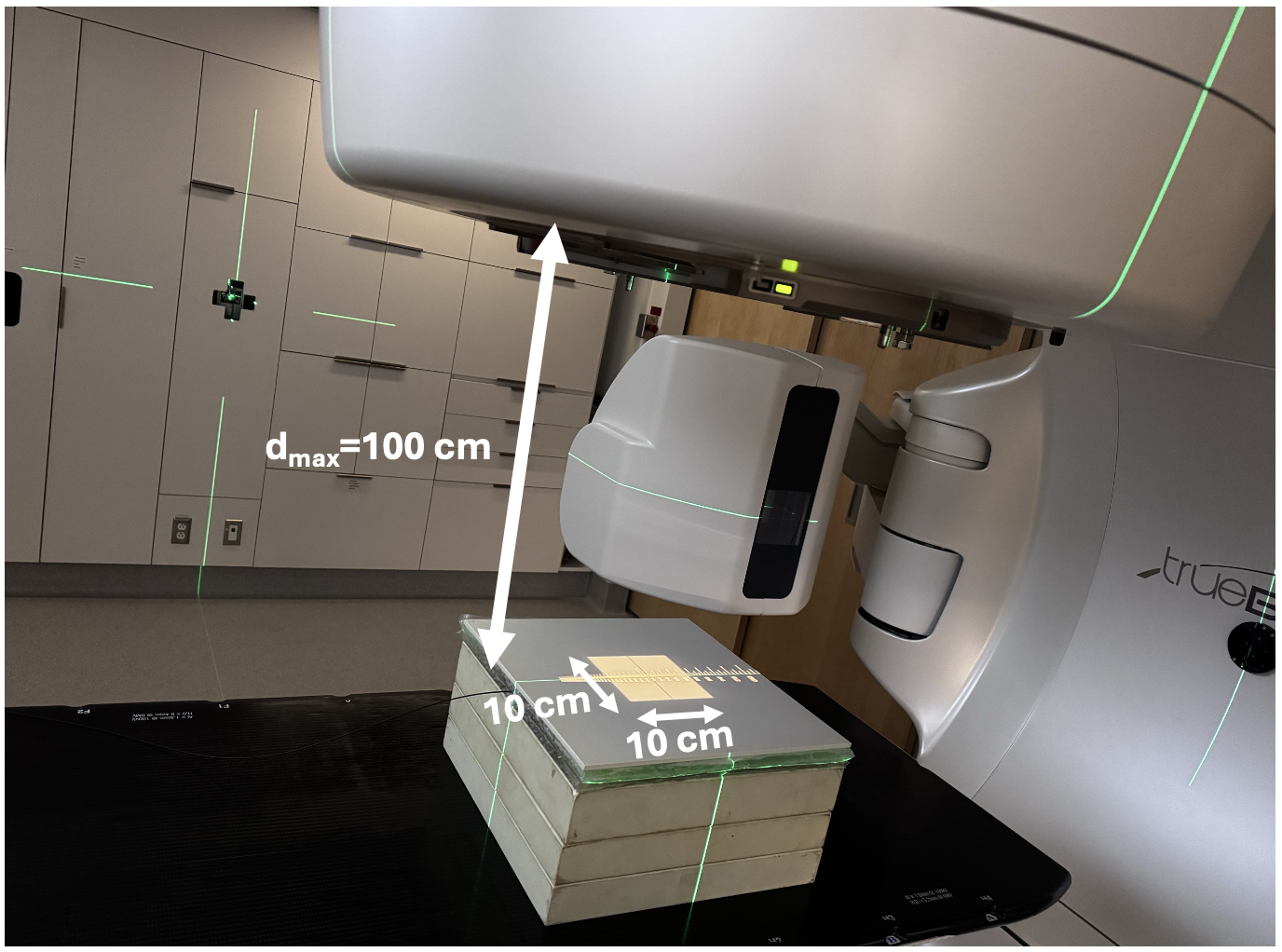}
    \caption{Experimental setup used for PSD dose calibration with a 6~MV beam. The tip was positioned at $d_{\text{max}} = 1.5$~cm depth in solid water at a Source to Detector Distance (SDD) of 100~cm, in the center of a $10 \times 10$~cm$^{2}$ reference field. A 500~MU irradiation was delivered to establish the conversion between scintillation light yield and absorbed dose to water.\\}
    \label{fig:DoseCalib}
\end{figure}

The detector calibration was performed based on the hyperspectral formalism \cite{Archambault2012}, enabling the independent removal of stem effects, according to the standard vendor guidelines at the linear accelerator \cite{Therriault2013, Gingras2024}. This was implemented to the HyperDose RP v1.5.3 software. In this work, a Varian TrueBeam Linac was used for both spectral and dose calibration.\\
For \textbf{spectral calibration}, measurements of the individual spectral components were performed, including scintillation, Cherenkov, fluorescence, and spectral attenuation of the clear fiber. The calibration steps (See Figure\ref{fig:SpecCalib}) are as follows:\\ 
\textit{Scintillation:} the Linac kV source was used with the following parameters: 90 kV, 154 mA, and 180 ms. Continuous mode was used without filters to obstruct the beam. The probe tip (PSD) was immobilized at the center of the output window  of the kV imager (See Figure\ref{fig:SpecCalib}a). The spectra was acquired during 60 s having a high Signal to Noise Ratio (SNR). \textit{Fluorescence:} the spectra was acquired with the Linac kV source using the same parameters as for scintillation spectra acquisition. 6 loops of the clear fiber of around 10 cm of diameter and about 30 cm away from the probe tip (See Figure\ref{fig:SpecCalib}b) were placed against the kV source and were irradiated 60 s reaching a high SNR. \textit{Cherenkov:} the spectra was extracted with the 6 MV beam at two gantry angles of 45° and 315°, to account for photon collection direction, as in one angle most photons are guided toward the spectrometer, while in the other they are guided toward the probe tip where they are absorbed. For each of these gantry angles, measurements were also taken at two positions on the clear fiber of 0.5 and 1.5 m from the tip to account for the fiber attenuation (See Figure\ref{fig:SpecCalib}c,d). The clear fiber was positioned at the isocenter on the top of solid water blocs to account for backscatter and was exposed to a $15\times15~cm$ field size being irradiated with 500 MU.\\
Finally for \textbf{dose calibration}, this was also performed with the 6 MV beam in reference conditions as follows (See Figure \ref{fig:DoseCalib}): the PSD tip was positioned at the middle of a 10x10 cm$^2$ field in solid water, at a depth of d$_{max}$=1.5 cm, and was irradiated with 500 MU to transform the collected light yield into absorbed dose which was 500 cGy in these reference conditions. All calibration measurements were acquired at a frame rate of 1 Hz (Standard mode).\\

\subsection{Brachytherapy measurements}
\begin{figure}[h!]
    \centering
    \includegraphics[width=1\textwidth]{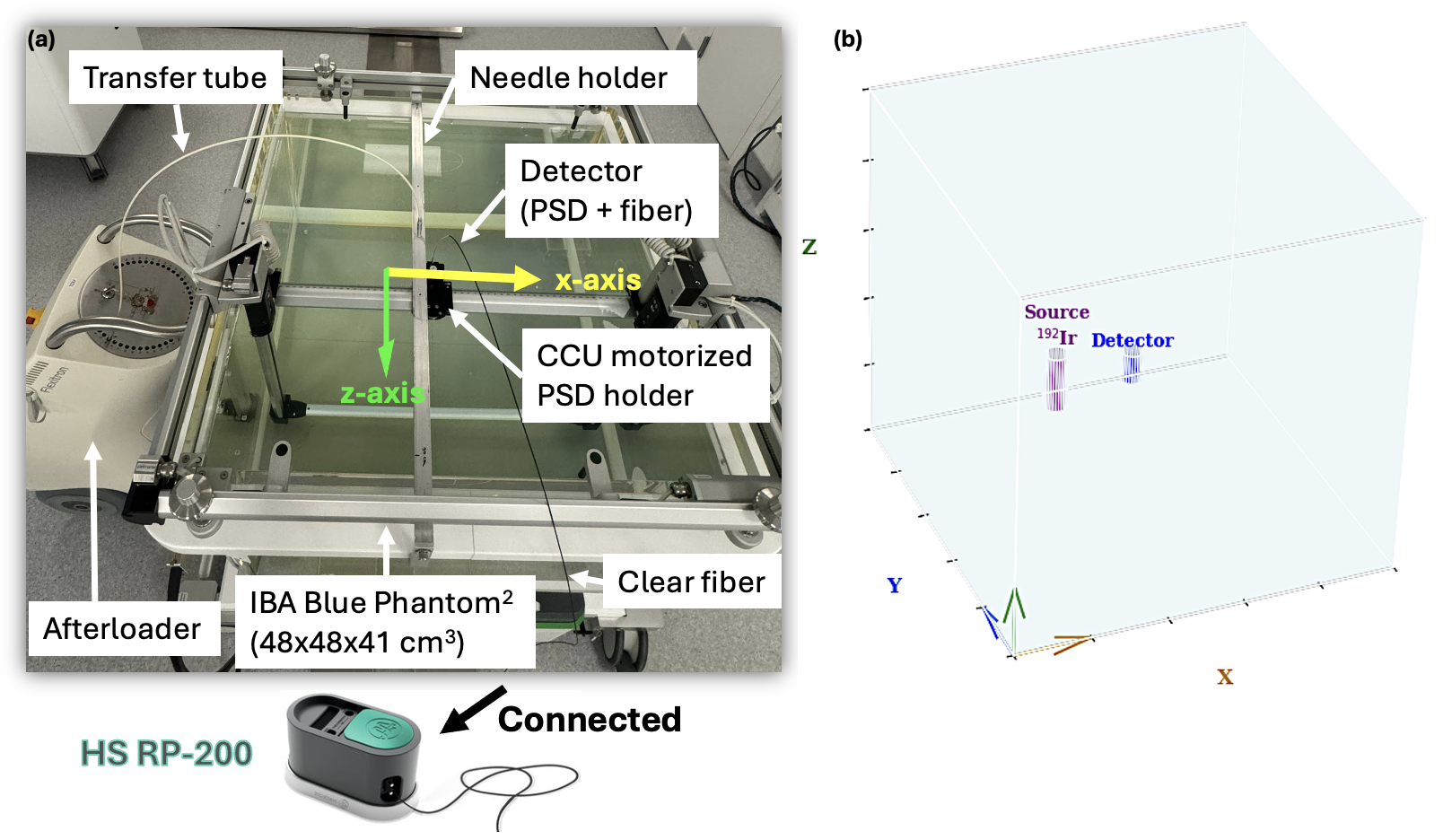} 
    \caption{Experimental setup at the BT unit for the MV-beam calibration validation. \textbf{(a)} brachytherapy measurement setup: the water tank from IBA showing the positioning unit of the PSD, catheter/source holder, and the Flexitron afterloader. \textbf{(b)} 3D representation of the geometry used in the experiment, showing the relative positioning of the \textsuperscript{192}Ir source and the PSD within the water phantom along with the displacement axes.\\}
    \label{fig:SetupExp}
\end{figure}

For calibration validation, all brachytherapy measurements were performed with a motorized IBA-Blue Phantom\textsuperscript{2} water tank whose dimension were 48×48×41 cm\(^3\), for compliance with TG-43U1 full scatter conditions (See Figure\ref{fig:SetupExp}). An Iridium-192 source (Flexisource, Elekta, Veenendaal, The Netherlands). Source-to-detector displacement, along the x-axis (source transverse plane, \(\theta=90^{\circ}\)) and the z-axis (source longitudinale plane), were computer-controlled with the IBA Common Control Unit (CCU) (Figure  \ref{fig:SetupExp}a).

After being calibrated at the linear accelerator as described above, the dosimetry system (HS RP200) was kept connected to an external power supply to maintain the stability of the reader’s electronic and temperature before being used for brachytherapy measurements. The Iridium-192 Flexisource with \(S_k=29447\) U was used. BT dose rate measurements were acquired along two axes. First, while the PSD was moving along the source transverse plane. The reference position in the x ($x_{ref}$), y ($y_{ref}$), and z ($z_{ref}$) axes was set as follows: in the x-axis, the PSD position was kept to around $1.2\pm0.05~cm$, distance between the two needles, one holding the source and the other holding the detector, and measured with a ruler. The reported uncertainty on the measured distance is based on the rule of the smallest division. This was the primary x reference position. Then, in the z direction, the PSD was moved in an interval around the source center till reaching the maximum dose or light yield; this was defined as the primary z reference position ($z_{ref}$).  Last, in the y direction once at the primary x and z reference positions, the PSD was also moved in an interval around the source till reaching the maximum. This was set as the primary y reference position ($y_{ref}$). In order to validate reference coordinates, calculated TG-43U1 dose rate at z and y=0 and x=1.2 cm ($\theta=90°$) was compared to the measured dose rate in our primer reference position. This confirmed well the reference coordinated to be ($x_{ref}=1.2\pm0.05~cm$, $y_{ref}=0~cm$, $z_{ref}=0~cm$ at the source center). It's of interest here to mention that uncertainties on TG-43U1 calculation are to take into account as it was used to fix the reference position. More details are in the uncertainty analysis section (see Section \ref{sec:uncertainty}). For measurements, the PSD position (x-axis) ranges from 1.2 to 8 cm (crossline scan by the CCU) (Figure ~\ref{fig:SetupExp}b). The second measurement set was while the PSD was moving along the z-axis of the source (depth scan by the CCU), the reference position was kept as defined earlier and the scan ranged from -4 to 8 cm around $x_{ref}$ and $z_{ref}$ position (Figure ~\ref{fig:SetupExp}b). Both sets of measurements were performed in 0.2~cm steps.

The accuracy of the calibration process, going from the linear accelerator to BT, was evaluated in terms of relative difference (RD) between measured ($\dot{D}_M$) and TG-43U1 calculated dose rates ($\dot{D}_{TPS}$) referred to as dose rates from the Treatment Planning System (TPS) with the corresponding $S_k$ value \cite{AAPM_TG43U1_2004}. The RD is calculated as follows:
\begin{equation}
    RD (\%) = \frac{\dot{D}_{TPS} - \dot{D}_M}{\dot{D}_{TPS}} \times 100
    \label{eq:RD}
\end{equation}
The TG-43U1 formalism was further used to convert RDs to positional errors \cite{Fonseca2020}, in order to quantify deviations in terms of possible detector displacement due to the CCU or IBA motorized unit and the afterloader reproducibility. To do so, the following equation from Johansen et al. and Fonseca et al. \cite{Johansen2018, Fonseca2020}, under the assumption of a point detector, is used to convert dosimetric differences into geometrical deviations, as the dose rate is inversely dependent on the square of source-to-dosimeter distance (SDD):
\begin{equation}
    \frac{\Delta\dot{D}}{\dot{D}} = \frac{-2}{r} \, \Delta r,
    \label{eq:deltarTG43}
\end{equation}
where, $\dot{D}$ is the dose rate, $\Delta\dot{D}$ is the dose rate deviation, $r$ is the dosimeter distance from the source or SDD, and $\Delta r$ is the geometric deviation.

During these measurements, the source position remained unchanged to eliminate the uncertainty in source delivery by the afterloader due to tube transfer curvature. However, afterloader reproducibility intervenes when the reproducibility of the reference position was explored by multiple deliveries of the source at that same planned dwell-position (see Section \ref{sec:uncertainty}). The acquisition frame rate was set to 10 Hz in this part of measurements.
\subsection{Uncertainty analysis} \label{sec:uncertainty}
To estimate the total uncertainty on the measured dose rates at any position \(r\), we consider the combined uncertainties arising from the detector response, the computer-controlled positioning unit, the afterloader, and the determination of the reference point according to TG-43U1 dose rates.
We first define the reference uncertainty derived from the reference parameters from the TG-43U1 formalism. This will be referred to as $u_{TPS}$ following the nomenclature from Andersen et al. \cite{Andersen2009}  and is defined as follows:
\begin{equation}
    u_{TPS}=\sqrt{u_{S_{K, CLINIC}}^2 + u_{MC}^2 + u_{TG43}^2}
    \label{eq:uTPS}
\end{equation}
where, \( u_{S_{K,CLINIC}}\) is the relative uncertainty on the source strength (source calibration in the clinic using the well-type ionization chamber in reference conditions \cite{DeWerd2011}), \( u_{MC} \) is the relative uncertainty in the dosimetry parameters of the source from Monte Carlo calculations defined by Dewerd et al. \cite{DeWerd2011} for the high energy source (Ir-192 Flexisource), \( u_{TG43} \) is the relative uncertainty related to interpolation methods used for dose calculation \cite{DeWerd2011}.

Next, we define the uncertainty that will be propagated from the calibration process of the reference point \( r_0 = 1.2 \, \text{cm} \), as follows:
\begin{equation}
u_{\text{ref}}(r_0) = \sqrt{u_{TPS}^2+ u_{\text{Afterloader}}^2},
\label{eq:reference_uncertainty}
\end{equation}
where, \( u_{\text{Afterloader}} \) is the reproducibility of the afterloader, delivering the source at a fixed position. This was assessed for 19 source transits by the afterloader to a fixed dwell-position at 200 mm on the check-ruler. The source center was then identified. 

The combined relative standard uncertainty $U_{c}(r)$ at any position \( r\) is then determined based on the law of propagation of uncertainties, the "root-sum-of-squares" method \cite{JCGMGUM, Taylor1994}, and is given by:
\begin{equation}
U_{c}(r) = \sqrt{u_{\text{ref}}^2+u_{\text{CCU}}^2+u_{M}^2} (r),
\label{eq:combined_uncertainty}
\end{equation}
where, \( u_{M} \) is the relative uncertainty on the mean measured dose rate at position \( r\) due to detector response/sensitivity, \( u_{\text{CCU}} \) is the reproducibility of the automatic positioning system (IBA automatic unit), and \( u_{\text{ref}} \) is as defined by Equation  \ref{eq:reference_uncertainty}. The uncertainty in the detector positioning by the CCU, \( u_{\text{CCU}} \), was evaluated by sending the automatic holder to a given position in the x-direction (see Figure \ref{fig:SetupExp}). Twenty (20) repeated measurements using a ruler were performed. For both CCU and afterloader, the standard deviation was calculated to evaluate the reproducibility. All uncertainty analyses in this work, reported and calculated, are within the 68\% confidence interval (k=1) unless otherwise specified.

In order to evaluate the statistical significance of deviations in dose rates when compared to a treatment planning system (or the TG-43U1 formalism), the statistical test proposed by Andersen et al. is adopted \cite{Andersen2009}. For a measured dose rate at each SDD or position r, the deviation is normalized to the combined uncertainty $U_c$ as estimated from our uncertainty budget above as follows \cite{Andersen2009}:
\begin{equation}
    \dot{D}_M-\dot{D}_{TPS}(r)= d\,U_{c}(r),
    \label{eq:deviation}
\end{equation}
where, d is the standardized deviation (or z-test) between the two values, measured and TPS\cite{Andersen2009}. This difference follows a normal or Gaussian distribution referred to as $N(\mu,\sigma^2)$ with a mean value of $\mu=0$ and a standard deviation of  $\sigma=1$. The probability of observing a deviation $d>2.58$ by chance is only of the level of 1 \%, corresponding to a p-value of 0.01. In other words, this deviation is no longer within the 99 \% confidence interval (CI) and is considered detectable. Similarly for 95 and 90 \% CIs, $d$ is $>1.96$ and $>1.64$ respectively. This deviation is, therefore, not explainable by the uncertainty budget alone but must also be attributed to \textit{another source of uncertainty} e.g. the detection of a shift in a dwell position due to a catheter shift or a reconstruction error. Thus, one could associate this \textit{other uncertainty} to positional displacement detection capability $\Delta r_c$, and is defined as follows according to the Equation \ref{eq:deltarTG43} as derived from the TG-43U1 formalism \cite{AAPM_TG43U1_2004}:
\begin{equation}
    \Delta r_c (r)=\frac{d}{2}\,r\,U_c(r).
    \label{eq:deltardetectable}
\end{equation}
We also define the ideal case where the detector response is excluded. Uncertainties are then only arising from combination of \( u_{\text{ref}} \) and \( u_{\text{CCU}} \) in the Equation \ref{eq:combined_uncertainty}. We define this uncertainty as: 
\begin{equation}
    u_{Pos}=\sqrt{u_{CCU}^2 + u_{Afterloader}^2+ u_{TPS}^2,}
    \label{eq:uPos}
\end{equation}
the positioning contribution from CCU and Afterloader to uncertainties combined to the TPS uncertainties. The ideal detector positional displacement detection capability $\Delta r_{Pos}$ is then defined as:
\begin{equation}
    \Delta r_{Pos} (r)=\frac{d}{2}\,r\,u_{Pos}(r).
    \label{eq:deltarideal}
\end{equation}

\section{Results}
\subsection{Dose rate measurements}\label{sec:doseratemeas}
\begin{figure}[h!]
    \centering
    \includegraphics[width=\textwidth]{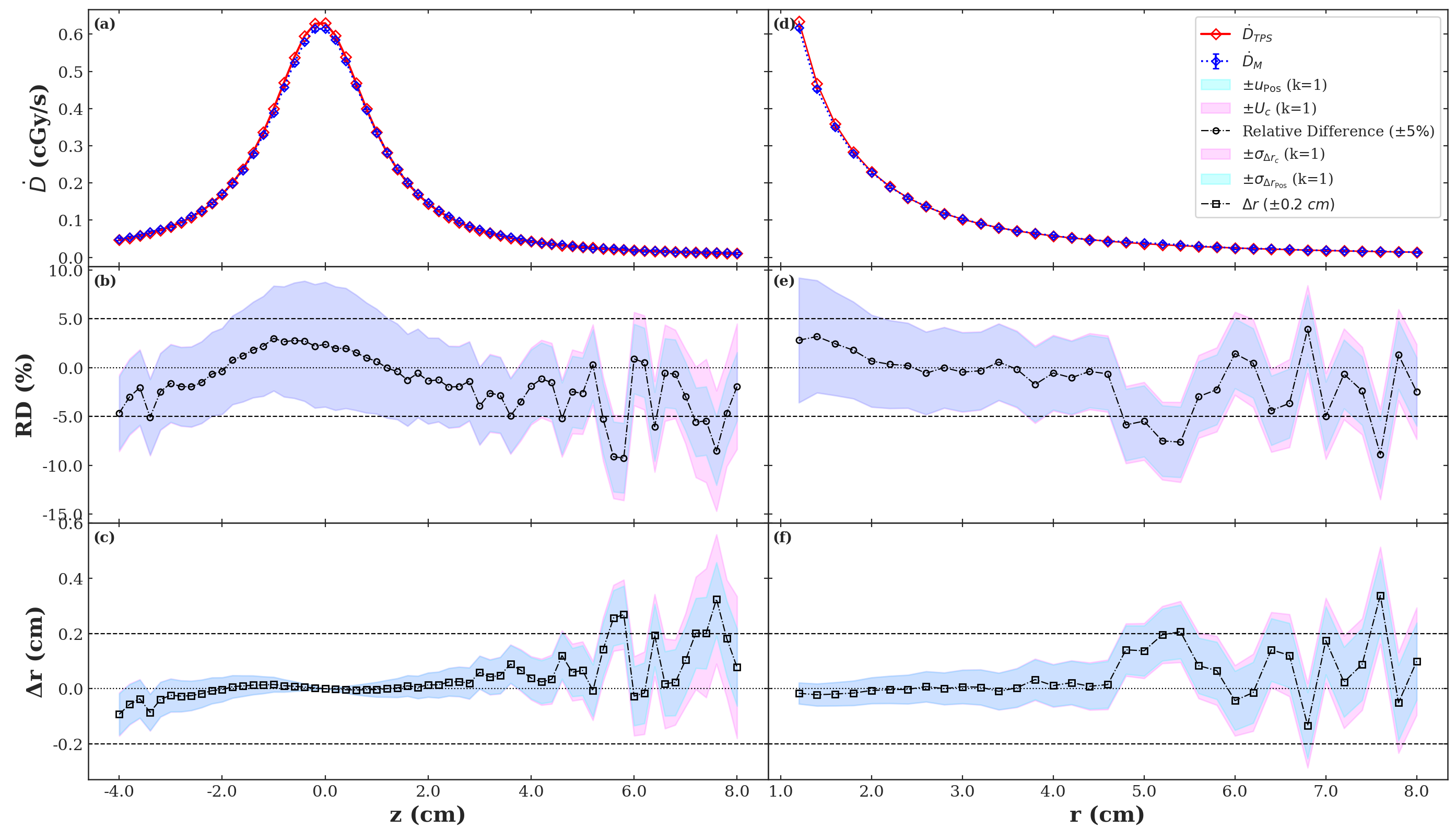}
    \caption{Comparison between measured dose rates ($\dot{D}_M$, blue) and TG-43U1‐based dose rates ($\dot{D}_{\text{TPS}}$, red) at 10~Hz acquisition rate. \textbf{(a)} Measured and calculated dose rates along the z-axis of the source (depth scan). \textbf{(d)} Measured and calculated dose rates along the x-axis within the source transverse plane at $\theta=90^{\circ}$ (crossline scan). Error bars on $\dot{D}_M$ correspond to the standard error of the mean ($k=1$) over 31 values per dwell-position frame. \textbf{(b)} and \textbf{(e)} show the relative difference RD between $\dot{D}_M$ and $\dot{D}_{\text{TPS}}$ (Equation~\ref{eq:RD}), with dotted lines indicating the $\pm 5~\%$ agreement range. For both b and e panel, the cyan band ($\pm~u_{Pos}$) represents the uncertainty on the RD from positional and TG43-U1 contribution only as defined in Equation \ref{eq:uPos}. The pink band represents the total uncertainty on the RD as propagated from Equation \ref{eq:RD} and as referred to in the uncertainty budget as $U_c$ from Equation\ref{eq:combined_uncertainty} including both $u_{\mathrm{Pos}}$ and the measurement uncertainty $u_M$. \textbf{(c)} and \textbf{(f)} panels show the positional error $\Delta r$  as calculated from Equation~\ref{eq:deltarTG43} including measurement and positional deviations in the RD values. The cyan band shows the propagated uncertainty $\sigma_{\Delta r,\mathrm{Pos}}$ due to $u_{\mathrm{Pos}}$ only, and the pink band the total propagated uncertainty $\sigma_{\Delta r,c}$ based on $U_c$. Dashed horizontal lines indicate the $\pm 0.2~\text{cm}$ positional tolerance recommended for clinical in vivo dosimetry \cite{Freniere2018, AAPM013}.}
    \label{fig:comparison}
\end{figure}

In both depth ($z$-axis) and crossline ($x$-axis) scans, results show a good agreement with RD values $\sim2.5\%$ at $r=1.2\pm0.05~cm$ from the source center This corresponds to a source-detector positional uncertainty, derived from Equation \ref{eq:deltarTG43}\cite{Fonseca2020}, of less than 0.15 mm (see Figure \ref{fig:comparison}). No correction factors were applied to the measured dose rates.

In Figure \ref{fig:comparison}, error bars in panels (b-e) were added by propagating the total uncertainty ($U_c$) through \ref{eq:RD} and as defined in the Equation \ref{eq:combined_uncertainty}. Uncertainties only due to TG43-U1 and positional errors as described by the Equation \ref{eq:uPos} were also added to highlight the contribution of measurement uncertainties (See Equation \ref{eq:uTPS}) due to detector sensitivity at close and large SDDs. The corresponding uncertainties on positional deviations were then propagated to $\Delta r$ values in panels (c-f) according to the Equation \ref{eq:deltarTG43}. Results show that in the z-axis (Figure  \ref{fig:comparison}a,b,c), at greater depths, RDs increase to more than 5\%, corresponding to positional uncertainty up to a maximum of 3 mm. The relative symmetry of the DR measurements along the z-axis further illustrates the excellent stem effect removal, which is maximized in the negative z-axis value and minimized in the positive z values. Along the x-axis (Figure  \ref{fig:comparison}d,e,f), RDs, around 2.5\% or less, are observed at distances close to the source, up to 4.6 cm. However, at greater SDDs, around 7.6 cm, the RD increases corresponding to a positional uncertainty between -0.15 and 0.30 cm.

\clearpage
In order to quantitatively evaluate the dose rates measured with the PSD, we assumed that uncertainty on the measured value at each corresponding SDD compared to the TG-43U1 values mainly arises from the PSD sensitivity, measurement equipment reproducibility, and the dosimetry parameters in the TG-43U1 calculations used for reference position calibration and for comparison. The following section (\ref{sec:uncertainty}) proposes an uncertainty budget de-convolving the contribution of each of these parameters to the uncertainty in our measurement to be faithfully compared to other detectors or references.

However, based on the uncertainty bands shown on the RD panels for both x and z scans, the total uncertainty on dose rate deviation crosses zero for $r > 3~\mathrm{cm}$ due to measurement uncertainties. Conversely, for $r < 3~\mathrm{cm}$, the error bars are higher or barely cross zero due to positional uncertainties being dominant in high gradient regions. This point is supported by the uncertainty bands in cyan due to only TG43-U1 and positional offset as described earlier, which is consistent with a dominant positional-offset contribution near the source being $\simeq$ to the total uncertainty ($U_c$) and decreasing at larger distances. This analysis is further confirmed in the following section.

\subsection{Uncertainties} \label{sec:uncertainty}
The absolute measured source positional uncertainty due to afterloader source delivery was 0.06 mm (k=1) at a fixed dwell position of 200 mm. This positional reproducibility ($u_{Afterloader}$) can be attributed to the afterloader motor accuracy and transfer tube curvature. CCU positional reproducibility  ($u_{CCU}$) was about 0.32 mm (k=1). Both absolute positional reproducibility ( $u_{Afterloader}$ and $u_{CCU}$) remain within $\pm$2 mm, in compliance with the CPQR (\textit{Canadian Partnership for Quality Radiotherapy}), and the AAPM guidelines \cite{Freniere2018, AAPM013}.
Table \ref{tab:uncertainty} summarizes the various sources of uncertainty that are propagated to our dose rate measurements related to both the Ir-192 Flexisource, the experimental set-up, and signal reading by our the detection system (PSD+HS-RP200).
\begingroup
\setstretch{1} % single spacing for the table
\renewcommand{\arraystretch}{1} % normal row spacing inside the table
\begin{table}[h!]
\centering
\small
\caption{Detailed uncertainty budget related to the experimental dose rate measurement setup in the brachytherapy unit}
\begin{tabularx}{\textwidth}{@{}p{4cm} c p{1.5cm} p{2.1cm} X@{}}
\toprule
\textbf{} & \textbf{Type} & \textbf{Symbol} & \textbf{Value} & \textbf{Comments/Estimation method} \\
\midrule
Source strength: $S_K$ (U) & B & $u_{S_{K, \mathrm{CLINIC}}}$ & $1.5\%$ & Uncertainty on the clinical measurement of the source air-kerma strength from the well-type ionization chamber during calibration (NIST-traceable) for high energy sources \cite{DeWerd2011}.\\
Monte Carlo dose estimate & B & $u_{MC}$ & $1.6\%$ & Uncertainty from MC simulations of dosimetry parameters including $u_{\Lambda}(\%)$, $u_{g(r)}(\%)$, $u_{F(r,\theta)}(\%)$, and $u_{\Phi_{an}(r)}(\%)$ for high energy sources \cite{DeWerd2011}.\\
Interpolation methods uncertainties & B & $u_{\mathrm{TG43}}$ & $2.6\%$ & Uncertainty from source and dosimetry data entered in the TPS for dose calculation using interpolation methods for high energy source \cite{DeWerd2011}.\\
Reproducibility of the Afterloader to the planned position according to the detector (SDD = 1.2 cm, absolute standard deviation with $k=1$) & A & $u_{\text{Afterloader}}$ & $0.059\ \mathrm{mm}$ & Reproducibility of the afterloader to the planned position using a caliper; the source center was identified on the check-ruler and the standard deviation over 19 repeated measurements was calculated. The relative uncertainty at $k=1$ is $0.03\%$ for a dose gradient of $16.7\%$ at the reference position of 1.2 cm from the source.\\
Detector sensitivity at position $r$ & A & $u_M$ & min: $0.15\%$, max: $3.39\%$ & At every position $r$ or SDD between $[-4, 8]$ cm, the detector was held stationary to acquire 31 repeated dose-rate readings. The relative standard deviation on the mean of these readings is then calculated, ranging between min and max values over the scanned distances.\\
Reproducibility of the automatic positioning system (absolute standard deviation) & A & $u_{\mathrm{CCU}}$ & $0.311\ \mathrm{mm}$ & Reproducibility of the CCU at a fixed position is here evaluated. The standard deviation was calculated over 20 repeated measurements. The relative uncertainty for $k=1$ is $0.106\%$ at a dose gradient of $16.7\%$ at 1.2 cm from the source.\\
\midrule
Combined uncertainty at the reference point ($k=1$) & A+B & $u_{\mathrm{ref}}$ & $3.4016\%$ & Equation~\ref{eq:reference_uncertainty}\\
Combined uncertainty at a position $r$ ($k=1$) & A+B & $U_c$ & min: $3.81\%$, max: $6.39\%$ & Equation~\ref{eq:combined_uncertainty}\\
\bottomrule
\end{tabularx}
\label{tab:uncertainty}
\end{table}
\endgroup
\clearpage
\subsubsection{Uncertainty analysis along the x-axis:}
In the following section we propose a detailed uncertainty analysis for the measured dose rates in the transverse plane of the source ($\theta=90°$) corresponding to the panels d, e, and f in the Figure \ref{fig:comparison} . 

\begin{figure}[htbp]
    \centering
    \includegraphics[width=\textwidth]{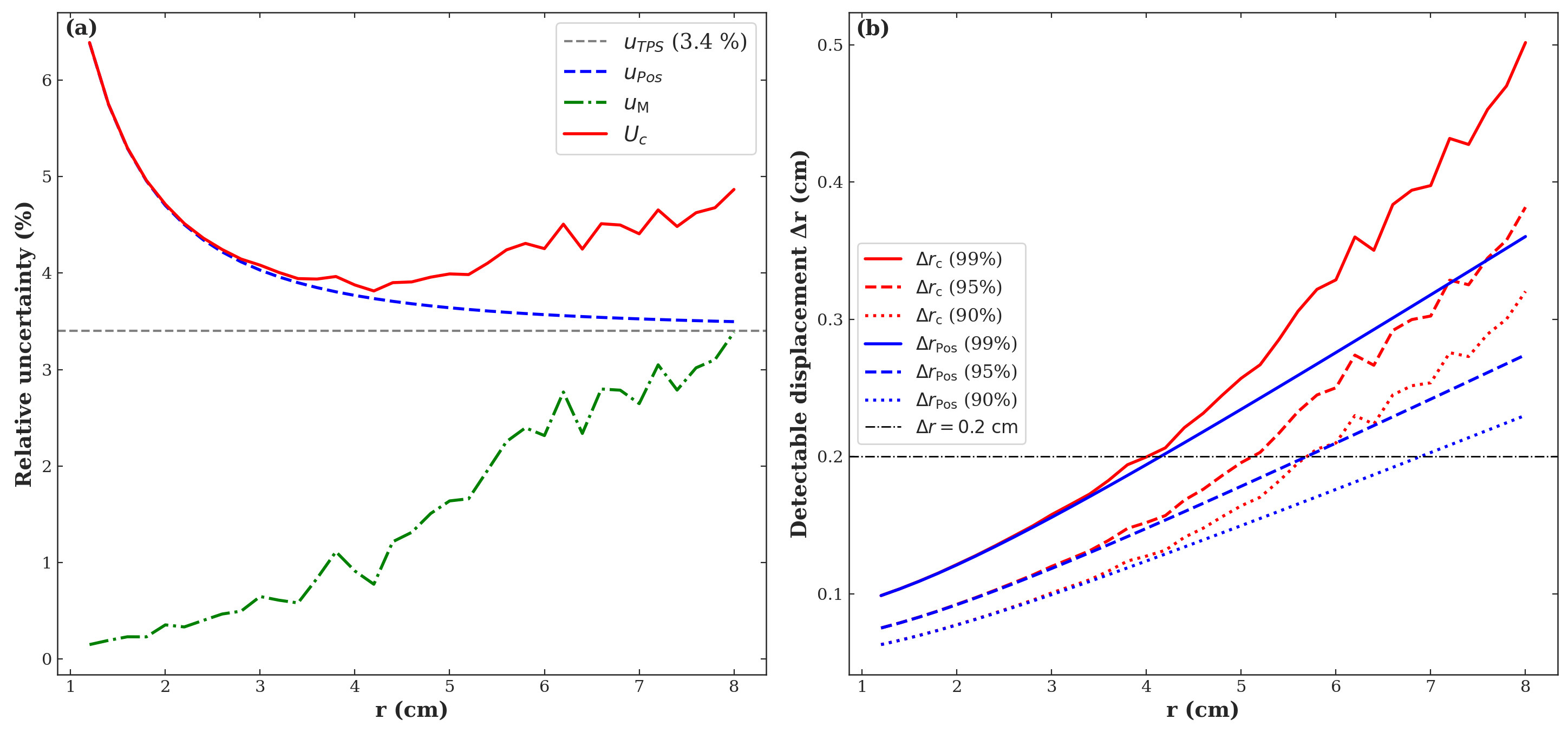}
    \caption{Relative dose rate uncertainties along the x-axis. \textbf{(a)} Uncertainties in dose rate measurements. Blue dashed line: the combined contribution of geometric positioning errors (CCU+Afterloader) converted into dose uncertainty plus the TG-138 uncertainty budget ($\sim 3.4\%$) in dashed gray line. Green dash-dot line: indicates the statistical uncertainty associated with the detector's measurement, or also the standard uncertainty on the mean (k=1). Red solid line: shows the total combined uncertainty, $U_{c}$.\textbf{ (b)} Shows the equivalent positional displacement due to $U_c$ in red and to the ideal case (excluding $u_{M}$) in blue at three different confidence intervals.\\}
    \label{fig:uncertainty}
\end{figure}

Figure \ref{fig:uncertainty} shows the different uncertainty components converted into relative dose rate uncertainties in percent at the corresponding SDDs in \textbf{(a)} (Equation  \ref{eq:deviation}) and the corresponding detectable displacement in \textbf{(b)} for three CI levels (Equation  \ref{eq:deltardetectable} and \ref{eq:deltarideal}). 
Dose rate measurements with the PSD were associated with a standard uncertainty on the mean denoted $u_{M}$ (k=1) by the (green) dash-dot lines (Figure  \ref{fig:uncertainty}a). The impact of positional uncertainties related to the afterloader and CCU reproducibility was evaluated at a fixed position and a fixed dose gradient, 16.7\%/mm at a distance of 1.2 cm. These positional, or mechanical, reproducibility were subsequently translated into dose rate uncertainties using the Equation \ref{eq:deltarTG43} \cite{Fonseca2020}, and were combined in quadrature with the uncertainties inherent to the TG-43U1 formalism \cite{DeWerd2011} as defined in the Equation \ref{eq:uPos}. $u_{Pos}$ is illustrated in Figure \ref{fig:uncertainty}a, by the dashed blue line. The total combined uncertainty $U_c$ (k=1) is shown by the red solid line in Figure \ref{fig:uncertainty}a, and is the quadrature sum of uncertainties arising from the propagated uncertainties due to TG-43U1 reference positioning, CCU/Afterloader reproducibility and detector's sensitivity as described in the Equation  \ref{eq:combined_uncertainty}. The highest $U_c$ is observed close to the source, in the high-dose gradient region, being dominated by the positional plus the reported TG-138 uncertainties of 3.4\% (k=1) for HDR brachytherapy  (Figure \ref{fig:uncertainty}a). This reaches a minimum value between 3 and 5 cm close to 3.4 \%, being the best that could be achieved even for a perfect dosimeter, and increasing progressively at larger distances to be dominated by the increased detector measurement uncertainty ($u_{M}$ in Figure \ref{fig:uncertainty}a).

In Figure \ref{fig:uncertainty}b, the corresponding detectable displacement calculated based on the TG-43U1 formalism \cite{Fonseca2020} as described in the Equation  \ref{eq:deltardetectable} and  \ref{eq:deltarideal}. The red curves correspond to the detectable displacement that can be measured taken into account the complete uncertainty chain (including the detector response, as given by Equation  \ref{eq:deltardetectable}), while the blue lines show the same quantity from uncertainties for an ideal dosimeter (Equation  \ref{eq:deltarideal}). Solid, dashed, and dotted lines correspond respectively to 99, 95, and 90 \% confidence interval (Figure  \ref{fig:uncertainty}b). The recommended maximum tolerance on applicator, catheter or needle positions of 2 mm is further provided by the horizontal black dash-dot line (Figure  \ref{fig:uncertainty}b) \cite{Fonseca2020, AAPM013}.
\clearpage

Uncertainty analysis results show that with the complete uncertainty chain ($U_c$), 2 mm displacements are detectable up to a SSD of approximately 4 cm, whereas displacements of at least 5 mm are needed to detect significant deviations at 8 cm from the source with 99\% CI ($\Delta r_c (99\%)$ in Figure  \ref{fig:uncertainty}b). For 95 and 90\% CIs, $\Delta r_c (95\%)$ and $\Delta r_c (90\%)$, 2 mm displacements are detectable up to 5 and 5.6 cm from the source, respectively (Figure  \ref{fig:uncertainty}b). In the \textit{ideal} detector scenario ($\Delta r_{Pos}$ in Figure~\ref{fig:uncertainty}b), TG-43U1 estimated uncertainties (3,4 \%)  plus positional uncertainties due to the CCU and the Afterloader limit the 2 mm detectable displacement, represented in black dash-dot line ($\Delta r=0.2$ cm), at SDDs of 4.2 cm (99\% CI), 5.8 cm (95\%), and 7 cm (90\% CI).
\section{Discussion}
A good agreement in dose rates between the measured and the TG-43U1 calculated values is observed in both x and z scans (Figure  \ref{fig:comparison}). RDs around the source center agreed within 5\% or less (or  \(\pm1\) mm or less) to approximately 5 cm from the source, in x and z-axis respectively. Observing the symmetry along the z-axis in Figure  \ref{fig:comparison}a is particularly interesting. In the negative values, the source is far from the clear collecting fiber, thus leading to little stem effect and the signal is due mainly to the light produced in the scintillator. In the positive z-axis values, the source is now moving along the clear fiber and away from the scintillating fiber, thus leading to conditions of increasing light from the stem effect, which is well corrected by the hyperspectral method using the parameters extracted from the proposed Linac calibration procedure. It further confirms the working hypothesis behind this calibration procedure of the relative energy independence in this energy range for PSDs, as expected from the ratio of $\mu_{en}/\rho$ and stopping power $S/\rho$ \cite{Beaulieu2016}.

At greater distances, Figure  \ref{fig:comparison}b and e, show that RD values increased up to 7.5\%, but stays within \(\pm2\) mm equivalent positional uncertainty (Equation  \ref{eq:deltarTG43}) almost up to 8 cm. In comparison, the study by Linares-Rosales et al. \cite{LinaresRosales2019} showed that the dosimetry system used was less effective beyond 6 cm due to a drop in dosimeter sensitivity, while the earlier study by Therriault-Proulx et al. \cite{Therriault2013} was limited to less than 3 cm because of constraints on light collection imposed by the detection geometry. Furthermore, for large SSD values, we attribute the increasing $\Delta r$ values (Figure  \ref{fig:comparison}c and f) to a decrease in the dosimetry system sensitivity (PSD and HyperScint RP200) 

The following discussion focuses on the detailed uncertainty analysis of our measurement set-up, evaluating both the impact of the positional reproducibility of the Afterloader and CCU and the detector sensitivity. The AAPM task group n°13 \cite{AAPM013} requires a positioning reproducibility of the source delivery to be less than $\pm$1 mm. The CPQR report \cite{Freniere2018} states that the source positioning tolerance should be within $\pm$2 mm for daily source position control and $\pm$1 mm accuracy for quarterly control and source change. In addition, recent studies on modern Afterloader have shown that dwell-position accuracy, including offsets or distal corrections, maintains an average deviation with all positions within 0.50 mm of planned values \cite{Bellezzo2019}. In this work, for afterloader reproducibility, we observed 0.06 mm (k=1) uncertainty in the source position, and 0.32 mm (k=1) for reproducibility of the robotic arm positioning. Both absolute standard uncertainties remain within the clinical tolerances mentioned above \cite{AAPM013, Freniere2018, Bellezzo2019}.

The impact of CCU/Afterloader positional accuracy decreases with increasing SDDs as shown in Figure \ref{fig:uncertainty}a, and the total uncertainty in dose rate difference becomes dominated by $u_{Pos}$ uncertainty (Figure  \ref{fig:uncertainty}a dashed blue line). This is directly due to the very large dose gradient in close proximity to the HDR source. For instance, at the reference SDD of $1.2\pm0.05~cm$, the dose rate gradient was approximately 16.7 $\%$/mm.  As we move away from the source, the dose gradient also falls and is approximately 6\%/mm at 3 cm. In contrast, the uncertainty in detector response, $u_{M}$ in Figure  \ref{fig:uncertainty}a), increases with distance reaching up to 3.4 \% at 8 cm. In low gradient regions, the impact of the positioning uncertainty is small (see $u_{Pos}$ in Figure  \ref{fig:uncertainty}a) and the main limitations is related to detector response at these greater distances (see $u_{M}$ in Figure  \ref{fig:uncertainty}a). 

The above impact aligns well with the uncertainty bands, total in pink and positional in cyan, in Figure \ref{fig:comparison}b,e. At small SDDs ($r<3~cm$) the positional uncertainty ($u_{Pos}$) almost overlaps the total uncertainty as propagated from the Equation \ref{eq:RD}, confirming that positional uncertainty from TG43-U1 and afterloader/CCU positioning is the governing factor in high-gradient regions. Also at larger SDDs $u_{Pos}$ decreases according to the total uncertainty remained higher, indicating that detector response uncertainty ($u_M$) dominates. This is supporting to the point mentioned above from the uncertainty analysis \ref{fig:uncertainty}.

This is further illustrated by the overall U-shaped behavior of the combined uncertainties $U_c$ (Figure  \ref{fig:uncertainty}a, full red line): near the source, the effect of the high dose gradient is dominating, and a small positioning uncertainty leads to large dose rate variation. Beyond 5 cm, the decreased light yield collection from the dosimeter is added significantly to the overall uncertainty (Figure  \ref{fig:uncertainty}a). $U_c$ decreases up to 3 cm from the source to reach a minimum uncertainty of around 4\% limited by the  3,4\% TG-43U1 uncertainties. This minimum $U_c$ value was in the range between 3 and 5 cm approximately (Figure  \ref{fig:uncertainty}a). Thus, based on our total uncertainty budget, we suggest that the best compromise between dose gradient and detector measurement uncertainty for a dose calibration done purely in BT conditions would be SDD between 3 and 5 cm.

With regard to using dose rate measurements to track the source position, as demonstrated by Johansen et al. \cite{Johansen2018} and discussed by Fonseca et al \cite{Fonseca2020}, Figure \ref{fig:uncertainty}b indicates that taking into account the complete uncertainty budget $U_c$ , a 1 mm displacement is detectable up to 1.8 cm, a 2 mm displacement up to 4.2 cm, and $\geq5$ mm at 8 cm within a very stringent 99\% CI. Therefore our extracted uncertainty chain still aligns well with source positional error detection tolerances required by the AAPM and the CPQR guidelines \cite{Freniere2018, AAPM013} up to 4 cm from the source (Figure  \ref{fig:uncertainty}b) at 99\% CI. In comparison,  Andersen et al. \cite{Andersen2009} reported that their system detects around 3 mm shifts near the source and limited to 16 mm or greater at an SDD of 5 cm using a 99\% CI. Therefore, our system shows higher sensitivity at greater distances. In addition, the uncertainty analysis shows that even a perfect detector in the same experimental setup and conditions would still be limited by the 3.4\%  uncertainty based on the TG-138\cite{DeWerd2011} and positioning errors ($u_{Pos}$), and is not able to identify sub‑millimeter displacements at any SDDs for 99\% CI, unless one relaxes the required CI. For example using instead 95\% CI, $u_{Pos}$ sets the 1 mm detectability threshold at up to $\approx2.2$ cm vs. $\approx3.0$ cm for 90\% CI (see $\Delta r_{Pos}$ in Figure  \ref{fig:uncertainty}b). More precisely, the detection of submillimeter displacements at short SDDs is not fundamentally limited by the detector sensitivity, but rather by the positional accuracy of the experimental setup, which in this work is already more precise than what could be achieved during a clinical procedure. Indeed, the uncertainty budget in Fig.~\ref{fig:uncertainty}a shows that, at short distances, the measurement uncertainty $u_{M}$ is very low, meaning that submillimeter shifts could theoretically be detected by the PSD. However, validation of this capability would require a positioning accuracy significantly below 1~mm. At large SDDs, the opposite situation arises: the positional uncertainty becomes negligible, but the detector response uncertainty $u_{M}$ increases due to lower dose rates and light yield fluctuations. In these low-gradient regions, measurement uncertainty dominates and submillimeter positional offsets can no longer be detected (Figure \ref{fig:uncertainty}b). 

As a final note, the standard 1 mm$^3$ PSD used in this study covers a viable SDD detection range for many clinical situations. Still, as highlighted earlier, the increased measurement uncertainty at large SDD is directly related to the reduced light yield (from the reduced dose rate) and thus the increased signal fluctuation during the acquisition. This reflects that the sensitivity of a 1 mm$^3$ PSD decreases at larger distances. In order to overcome this issue, increasing the PSD volume to 2 or even 3 mm$^3$ keeping the same diameter would result in a similar increased light yield by a factor of 2 or 3, and reduced uncertainties at these large distances if needed for clinical applications. Future work will concentrate on extracting dwell-times from the measured dose rate profiles and confirming the ability of the dosimetry system to detect delivery errors. This would be feasible by simulating geometric or positional displacement errors, like guide-tube interchange and applicator movement. The statistic test will be applied to quantify the system’s sensitivity in terms of deviation threshold \cite{Andersen2009}.
\section{Conclusion}
The results presented in this study validate the effectiveness and robustness of a 6 MV external beam calibration method, in well-controlled standard conditions, based on the hyperspectral technique for a PSD to be used for time-resolved HDR brachytherapy measurements. We note that the proposed calibration method is generic and can be applied to a single and multi-point plastic scintillation detector, even though it was demonstrated here using a specific commercial model. Uncertainties close to the source are consistent with the reproducibility associated with the afterloader and IBA motorized units. According to the uncertainty analysis, positional uncertainty near the source dominates due to the dose gradient, whereas the detector response dominates at greater distances. The detailed uncertainty budget further demonstrates the intrinsic limitation in our ability to detect small changes in source positions using IVD for source tracking.
\section*{Acknowledgments}
Authors gratefully acknowledge the Medscint team for their support in the design and construction of the PSD used in these experiments. We also thank Dr. Luc Gingras for his valuable contribution to the establishment of the uncertainty budget.

This research was enabled in part by the financial support from the Natural Sciences and Engineering Research Council of Canada (NSERC) Discovery grants (RGPIN 2019-05038) held by Pr. Luc Beaulieu.
\section*{Data availability statement}
The data supporting this study can be obtained from the authors upon reasonable request.
\newpage
\section*{References}
\addcontentsline{toc}{section}{\numberline{}References}
\vspace*{-20mm}
\bibliography{Bib}
\bibliographystyle{medphy}

\end{document}